\documentclass[accepted]{uai2023} 

\usepackage[american]{babel}
\usepackage{colortbl}
\usepackage{tabularray}

\usepackage{natbib} 
\usepackage{mathtools} 
\usepackage{booktabs} 
\usepackage{tikz} 
\usepackage{dsfont}
\usepackage{amsmath,amssymb}
\usepackage{subfig}

\newcommand{\our}{CrysMMNet}

\title{CrysMMNet: Multimodal Representation for Crystal Property Prediction}

\author[1]{\href{mailto:kishalaydas@kgpain.iitkgp.ac.in}{Kishalay Das}{}}
\author[1]{Pawan Goyal}
\author[2]{Seung-Cheol Lee}
\author[2]{Satadeep Bhattacharjee}
\author[1]{Niloy Ganguly}
\affil[1]{%
    Department of Computer Science \& Engineering\\
    Indian Institute of Technology, Kharagpur, India\\
}
\affil[2]{%
    Indo Korea Science and Technology Center\\
    Bangalore, India\\
    }
  
\begin{document}
\maketitle
\begin{abstract}
   Machine Learning models have emerged as a powerful tool for fast and accurate prediction of different crystalline properties. Exiting state-of-the-art models rely on a single modality of crystal data i.e crystal graph structure, where they construct multi-graph by establishing edges between nearby atoms in 3D space and apply GNN to learn materials representation. Thereby, they encode local chemical semantics around the atoms successfully but fail to capture important global periodic structural information like space group number, crystal symmetry, rotational information etc, which influence different crystal properties.  In this work, we leverage textual descriptions of materials to model global structural information into graph structure and learn a more robust and enriched representation of crystalline materials. To this effect, we first curate a textual dataset for crystalline material databases containing descriptions of each material. Further, we propose \our{}, a simple multi-modal framework, which fuses both structural and textual representation together to generate a joint multimodal representation of crystalline materials. We conduct extensive experiments on two benchmark datasets across ten different properties to show that \our{} outperforms existing state-of-the-art baseline methods with a good margin. We also observe that fusing the textual representation with crystal graph structure provides consistent improvement for all the SOTA GNN models compared to their own vanilla versions. We have shared the textual dataset, that we have curated for both the benchmark material databases, with the community for future use. \\
\end{abstract}
\section{Introduction}
\label{intro}
In the recent past, we have witnessed a surge of interest in developing machine learning models ~\cite{seko2015prediction,pilania2015structure,lee2016prediction,de2016statistical,seko2017representation,isayev2017universal,ward2017including,lu2018accelerated,im2019identifying} for fast and accurate property prediction of crystalline materials. 
Crystalline materials are typically modeled by a minimal unit cell containing all the constituent atoms in different coordinates, repeated infinite times in 3D space on a regular lattice, which makes material structures periodic in nature. A key challenge in learning crystal representation is how to capture accurately global periodic structural information along with local chemical semantics. Recent state-of-the-art models ~\cite{xie2018crystal,chen2019graph,louis2020graph, Wolverton2020,schmidt2021crystal,choudhary2021atomistic,hsu2021efficient,das2022crysxpp,yan2022periodic} construct multi-edge graphs for a 3D material structure where they create edges between nearby atoms within a pre-specified distance threshold in 3D space and apply GNN model to learn representations of crystal structures that are optimized for downstream property prediction tasks. Although existing variants of GNN models predict different crystal properties with high precision, they rely on a single modality of crystal data i.e crystal graph structure which limits the expressive power of these models. The architectural innovations of these approaches are primarily based on incorporating specific domain knowledge of the local bonding environment, such as explicitly encoding bond angle ~\cite{choudhary2021atomistic}, dihedral angle ~\cite{hsu2021efficient}, etc. but they fail to incorporate crucial global periodic structural information like lattice constraint, space group number, crystal symmetry, rotational information, component 3D orientation, heterostructure information, etc, which will enrich its representation and subsequently aid the property prediction accuracy.\\
 In this work, we propose to learn a more robust and enriched representation by using multi-modal data i.e graph structure and textual description of materials. One of the major advantages of using the textual description of materials is it provides a diverse set of periodic structural information which is useful to estimate different crystal properties but difficult to incorporate explicitly into a graph structure. Leveraging textual modalities beyond graph structures of materials remains unexplored by the research community and to the best of our knowledge, there is no existing dataset containing textual descriptions of the materials. Hence, we first curate the textual dataset of two popular materials databases (Graph-based), Material Project (MP) and JARVIS, containing textual descriptions of each material of those databases. We used a popular tool robocrystallographer ~\cite{ganose2019robocrystallographer} to generate descriptions for global crystal structures, which looks at the structural symmetry, local environment, and extended connectivity to generate a description that includes space group number, crystal symmetry, rotational information, component orientations, heterostructure information, etc.\\
 Further, we propose, \our{} (\textbf{Crys}tal \textbf{M}ulti-\textbf{M}odal \textbf{Net}work), a simple multi-modal framework for crystalline materials, which has two components: Graph Encoder and Text Encoder. Given a material, Graph Encoder uses its graph structure and applies GNN based approach to encode the local neighborhood structural information around a node (atom), and subsequently learn graph (crystal) representation. On the contrary, Text Encoder is a transformer-based model, which encodes the global structural knowledge from the textual description of the material and generates a textual representation. Finally, both graph structural and textual representation are fused together to generate a more enriched multimodal representation of materials, which captures both global and local structural knowledge and subsequently improves property prediction accuracy.\\
 To show the merit of our proposed algorithm, we performed comprehensive experiments on two popular benchmark datasets, Materials Project and JARVIS-DFT, across ten diverse sets of properties and compare the results with popular state-of-the-art models. We observe that for all the properties \our{} can achieve the lowest error in comparison with other baseline models. In addition, our results demonstrate that multi-modal representation learning  helps to achieve even better improvements when the dataset is sparse. We also perform some ablation studies to investigate the expressiveness of textual representation and robustness of multimodal representation on different GNN architectural choices. Result shows, textual representations alone are not expressive enough to learn the structure-property relationship of the materials.
 Moreover, fusing both graph structural and textual representation together leads to substantial performance improvements for all the state-of-the-art GNN models compared to their vanilla versions. We also investigate the influence of local compositional information and global material structural knowledge encoded through
textual representation and found for all crystal properties both local and global knowledge improves the downstream property prediction accuracy. 
We have shared the textual dataset, that we have curated for both the benchmark material databases with the community for future use.\footnote{Source code and dataset of \our{} is made available at \url{https://github.com/kdmsit/crysmmnet}}
\begin{figure*}
	\centering
	\includegraphics[width=1.9\columnwidth]{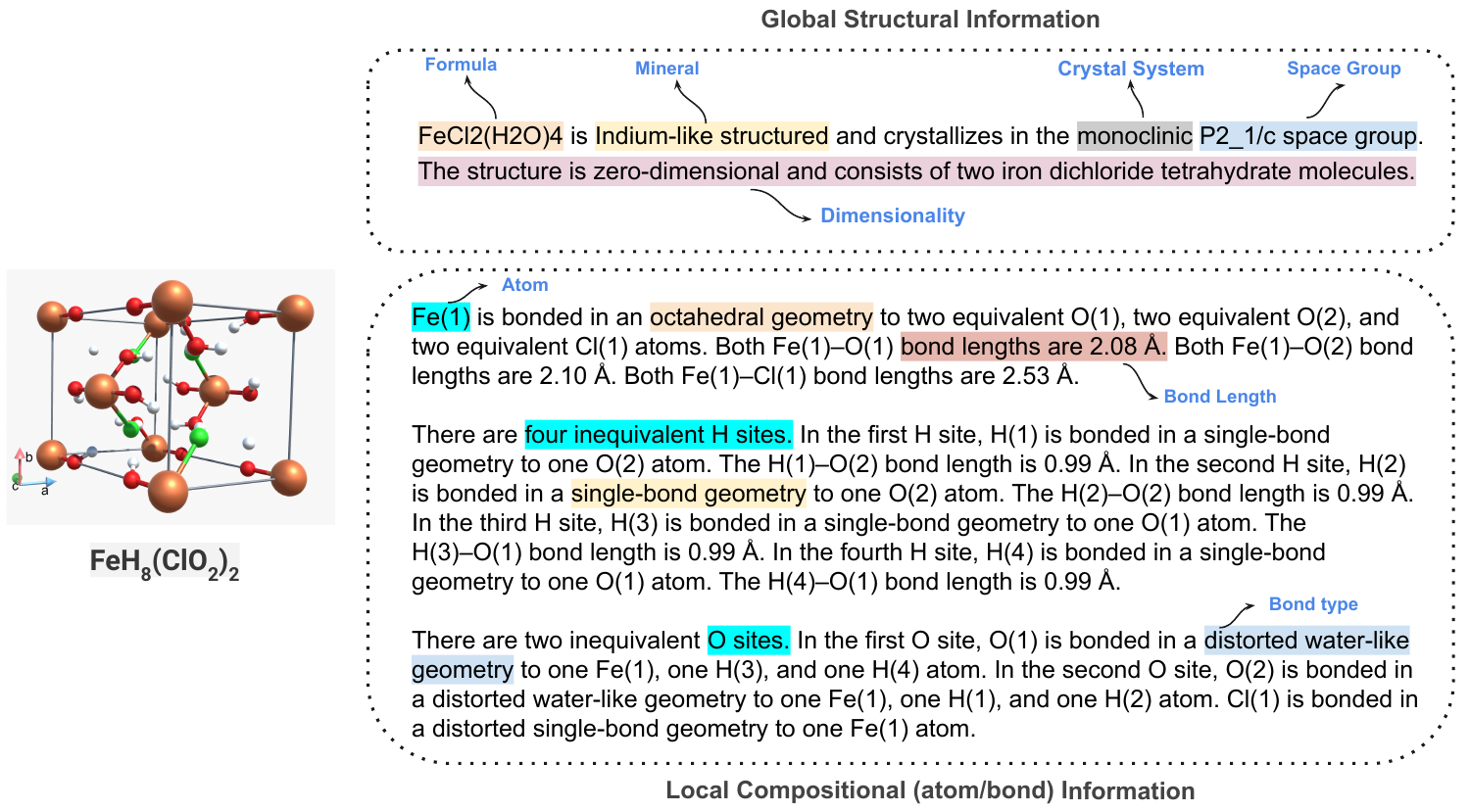}
	\caption{Textual description of $\mathbf{FeH_8(ClO_2)_2}$ material from JARVIS dataset generated by Robocrystallographer. The generated text contains both local chemical compositional information related to  atom/bonds (like site coordination, geometry, polyhedral connectivity, and tilt angles) and global structural knowledge (like mineral type, space group information, symmetry, and dimensionality).}
	\label{fig:text}
\end{figure*}
\section{Background and Related Work}
\subsection{Crystal Representation}
The structure of a crystalline material can be modeled by a minimum unit cell, repeated infinite times in three-dimensional (3D) euclidean space on a regular lattice, which makes the crystalline structure periodic in nature. As mentioned in ~\cite{xie2021crystal,yan2022periodic}, for a given crystal we can describe its unit cell by two matrices: Feature Matrix ($\mathbf{X}$) and Coordinate Matrix ($\mathbf{C}$). Feature Matrix $\mathbf{X} =[\mathbf{x}_1,\mathbf{x}_2,...,\mathbf{x}_n]^T \in R^{n \times d}$ denotes atomic feature set of the material, where $\mathbf{x}_i \in R^{d}$ corresponds to the d-dimensional feature vector of i-th atom. On the other hand, Coordinate Matrix $\mathbf{C}=[\mathbf{c}_1,\mathbf{c}_2,...,\mathbf{c}_n]^T \in R^{n \times 3}$ denotes atomic coordinate positions, where $\mathbf{c}_i \in R^{3}$ corresponds to cartesian coordinates of i-th atom in the unit cell. Further, there is an additional lattice matrix $\mathbf{L}=[\mathbf{l}_1,\mathbf{l}_2,\mathbf{l}_3]^T \in R^{3 \times 3}$, which describes how a unit cell repeats itself in the 3D space towards $\mathbf{l}_1,\mathbf{l}_2$ and $\mathbf{l}_3$ direction to form the periodic 3D structure of the material. Formally, a given crystal can be defined as $\mathbf{M}=(\mathbf{X},\mathbf{C},\mathbf{L})$ and we can represent its infinite periodic structure as 
\begin{equation}
    \label{eq:crystal}
    \begin{split}
    & \mathbf{\hat{C}}  = \{ \mathbf{\hat{c}}_i |  \mathbf{\hat{c}}_i = \mathbf{c}_i + \sum_{j=1}^{3} k_j\mathbf{l}_j \}; \:
    \mathbf{\hat{X}}  = \{ \mathbf{\hat{x}}_i |  \mathbf{\hat{x}}_i = \mathbf{x}_i \}
    \end{split}
\end{equation}
where $k_1,k_2,k_3, i \in Z, 1 \leq i \leq n$.

\subsection{Crystal Property Prediction using GNNs}
Graph neural networks have emerged as highly promising models in various domains of computer science, showcasing significant potential in many real-world applications including social networks \cite{hamilton2017inductive,chen2018fastgcn,dai2018learning}, recommender systems \cite{berg2017graph,ying2018graph}, hyper-networks \cite{yadati2019hypergcn,bandyopadhyay2020hypergraph}, chemical and biological networks \cite{duvenaud2015convolutional,gilmer2017neural} etc.
Recently, graph neural network (GNN) based approaches have been very effective to encode  structural information of the crystal materials into enriched embedding space so that it can predict different crystal properties with high accuracy. CGCNN \cite{xie2018crystal} is the first proposed model, which represents 3D crystal structure as an undirected weighted  multi-edge graph $\mathcal{G} =(\mathcal{V}, \mathcal{E}, \mathcal{X}, \mathcal{F})$ where $\mathcal{V}$ denotes the set of nodes (atoms) in the unit cell of material and $\mathcal{E}=\{(u,v,k_{uv})\}$ denotes a multi-set of node pairs and $k_{uv}$ denotes number of edges between a node pair $(u,v)$. $\mathcal{X}=\{(x_{u} | u \in \mathcal{V})\}$ denotes the node feature set, which includes different chemical properties like electronegativity, valance electron, covalent radius, etc. Finally, $\mathcal{F}_i=\{\{s^k\}_{(u,v)} |  (u,v) \in \mathcal{E}, k\in\{1..k_{uv}\}\}$ denotes the multi-set of edge weights where $s^k$ corresponds to the $k^{th}$ bond length between a node pair $(u,v)$, which signifies the inter-atomic bond distance between two atoms. Further, CGCNN develops a graph convolution neural network to update node features based on their local chemical and structural environment. \\
Following CGCNN, there are a lot of subsequent studies ~\cite{chen2019graph,louis2020graph, Wolverton2020,schmidt2021crystal}, where authors proposed different variants of GNN architectures for effective crystal representation learning. Through multiple layers of graph convolutions, these models can implicitly encode many-body interactions. Further, ALIGNN ~\cite{choudhary2021atomistic} explicitly captures many-body interactions by incorporating bond angles and local geometric distortions into the GNN encoding module to enhance property prediction accuracy and became SOTA for a large range of properties. 
Recently, transformer-based architecture Matformer ~\cite{yan2022periodic} is proposed to learn the periodic graph representation of the material, which is invariant to periodicity and can capture repeating patterns explicitly. Matformer marginally improves the performance compared to ALIGNN, however, is much faster than it.
Moreover, scarcity of labeled data makes these models difficult to train for all the properties, and recently, some key studies \cite{jha2019enhancing,das2022crysxpp,das2023crysgnn} have shown promising results to mitigate this issue using transfer learning, pre-training, and knowledge distillation respectively.
\section{Methodology}
\label{methodology}
\begin{figure*}
	\centering
        \includegraphics[width=1.9\columnwidth]{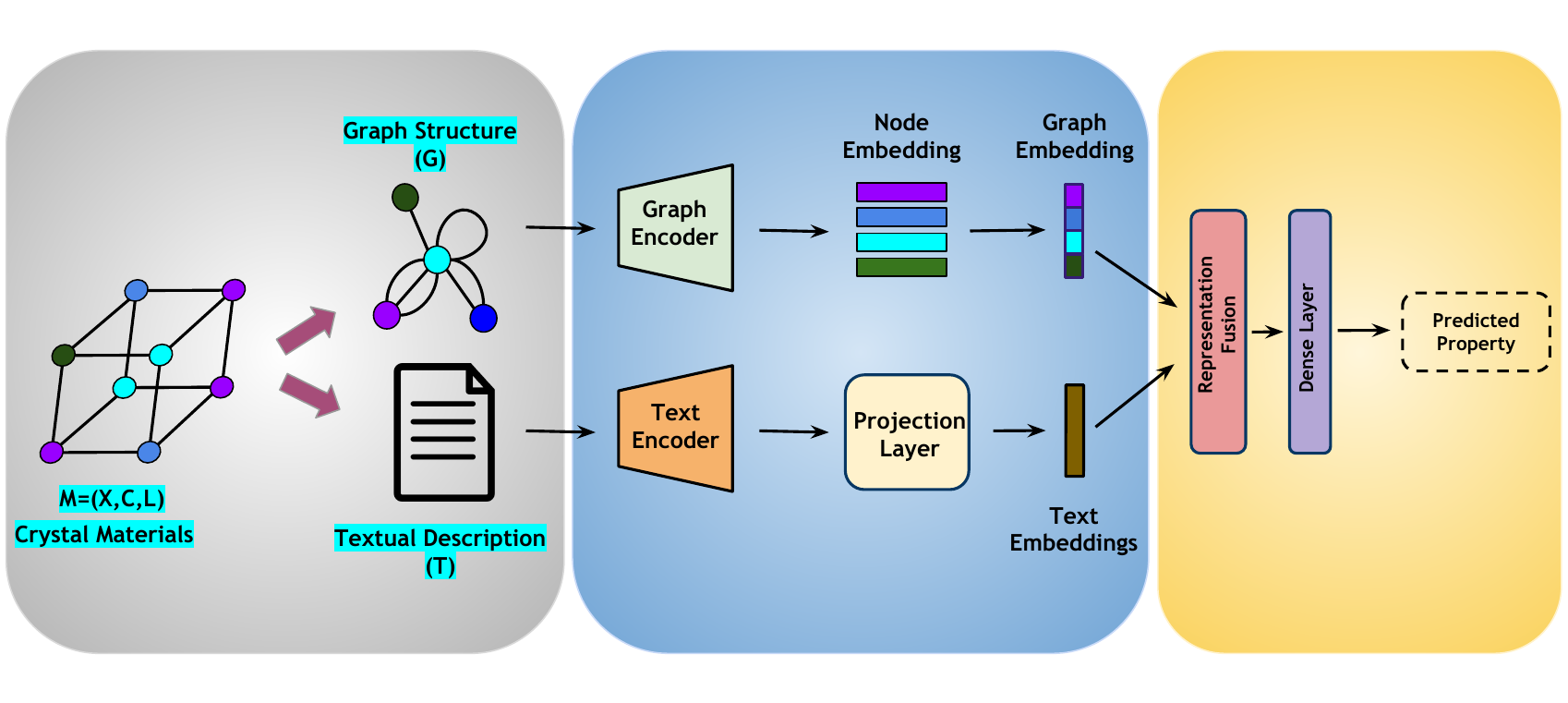}
	\caption{Overview of our adopted methodology \our{}. Given Crystal Material (M), we use two modalities - Graph Structure $(\mathcal{G})$ and Textual Description $(\mathcal{T})$. Graph structure $(\mathcal{G})$ is passed through a graph encoder to generate graph embedding $(\mathcal{Z}_\mathcal{G})$. Textual Description $(\mathcal{T})$ is fed through a text encoder followed by a projection layer to generate text embedding $(\mathcal{Z}_\mathcal{T})$. Finally, both the representations are fused together and predict the crystal properties based on joint modeling of the input modalities.}
	\label{fig:crysmmnet}
\end{figure*}
In this section, we first discuss the insights on the textual dataset that we have curated for two popular crystalline databases and explain the local compositional and global periodic information we are able to encode using textual representation, which is difficult to incorporate explicitly into a graph structure. Then we give a detailed overview of our proposed muti-modal framework \our{} that generates joint embedding for materials, which facilitates accurate property prediction. 
\subsection{Textual Dataset}
\label{textual}
Leveraging textual modalities beyond the conventional graph structure of the materials to capture both local atomic and global periodic knowledge remains largely unexplored by the research community. To the best of our knowledge, there is no existing dataset containing textual descriptions of the materials. Hence, we first curate the textual dataset for two popular material databases JARVIS and Material Project(MP), containing textual descriptions for each material of those databases. Conventionally, in these databases, the periodic structure of the materials is represented in Crystallographic Information File (CIF File). We use Robocrystallographer~\cite{ganose2019robocrystallographer}, which is a free utility, to generate a textual description of the material from the CIF file. Robocrystallographer decomposes crystal structures into local compositional (site coordination, geometry, polyhedral connectivity, and tilt angles) and global structural (mineral type, space group information, symmetry, dimensionality) components (Figure \ref{fig:text}) and output this information in three formats: JSON for machine use, human-readable text, and machine learning format. In this work, we use human-readable text for collecting textual datasets, which are easily interpretable and resembles a human description of the crystal structure.\\
\textbf{Local compositional} information describes local chemical environments around different atoms and inter-atomic bonds in a unit cell. It provides a detailed description of different sites of the materials, like atomic compositions of different sites, site coordination, inter-atomic connectivity through chemical bonds, bond type, and length. Further, the geometry of each site is mentioned and the presence of corner-sharing tetrahedra connectivity is specified. On the contrary, \textbf{global structural} information illustrates the global environment i.e. periodic structure and orientation of the material in 3D space. The most useful information it provides is regarding crystal symmetry, which includes the specific space group  and crystal system the material belongs to. Space group is used to describe the symmetry of a unit cell of the crystal material in 3D space. In materials science literature there are 230 unique space groups and each crystal (graph) has a unique space group number. Further based on the space group level information can classify a crystal graph into 7 broad groups of crystal systems like Triclinic, Monoclinic, Orthorhombic, Tetragonal, Trigonal, Hexagonal, and Cubic. Moreover, it contains the mineral type of the material and the dimensionality of the crystal structure. \\
Minerals are naturally occurring, inorganic substances with a specific chemical composition and a crystalline structure. The most common types include silicates (which contain silicon and oxygen), carbonates (which contain carbon and oxygen), sulfates (which contain sulfur and oxygen), halides (which contain a halogen element), oxides (which contain oxygen and one or more other elements), and sulfides (which contain sulfur and one or more other elements). Examples of minerals in each category include quartz, calcite, gypsum, halite, hematite, and pyrite. The chemical composition and crystal structure of a mineral determine its properties, such as its hardness, color, and cleavage.
Further, dimensionality of a material is a significant global feature that refers to the number of dimensions that a particular component of the material spans. The dimensionality of a bonded cluster of atoms can be determined by calculating the rank of the subspace spanned by the central atom and its periodically connected neighbors.\\
A comprehensive understanding of both local and global environments is necessary for robust prediction of material properties. For example, in the case of formation energy, the local chemical environment, such as atom composition, bond length, and bond angles, plays a crucial role in determining the electronic and geometric structure of the material, which directly affects its formation energy. A slight variation in the local environment can result in significant changes in the electron density and, subsequently, the energy required to form the materials. Similarly, the global chemical environment, like the space group, has a profound impact on the formation energy by controlling the arrangement of atoms within the material. Different space groups are associated with different crystal structures and packing arrangements, which can lead to different formation energies.  Moreover, the study by Larsen \textit{et. al}~\cite{Larsen} showed that the formation energies of layered materials can be related to their dimensionality, highlighting the importance of considering this feature in the investigation of materials.
\subsection{Multi-Modal Framework}
Next, we propose a simple, yet effective multi-modal framework, \our{}, for graph and textual embedding of materials, which realizes material dataset  as $D =\{ (\mathcal{G}, \mathcal{T}),\mathcal{Y} \}$, where $\mathcal{G}$,$\mathcal{T}$ and $\mathcal{Y}$ denote multi-graph structure, textual description and property value the material respectively.  In our multi-modal architecture, the goal is to learn a
function $f_{\theta}(\mathcal{G}, \mathcal{T})$
\begin{equation}
    \label{eq:func}
    \begin{split}
    f_{\theta} : (\mathcal{G}, \mathcal{T}) \rightarrow \mathcal{Y}
    \end{split}
\end{equation}
By design, \our{} (as shown in Figure \ref{fig:crysmmnet}) is composed of three modules: graph encoder $M_V(\mathcal{G}) \rightarrow \mathcal{Z}_{\mathcal{G}}$, text encoder $M_L(\mathcal{T}) \rightarrow \mathcal{Z}_{\mathcal{T}}$, and joint embedding model $E(\mathcal{Z}_{\mathcal{G}},\mathcal{Z}_{\mathcal{T}}) \rightarrow \mathcal{Z}$, where $\mathcal{Z}_\mathcal{G}$,$\mathcal{Z}_\mathcal{G}$ and $\mathcal{Z}$ are graph-level, textual and multimodal embedding respectively. Next, we explain each part of the \our{} framework in detail. 
\subsubsection{Graph Encoder: }\our{} adopts a GNN architecture inspired by ALIGNN \cite{choudhary2021atomistic} as Graph Encoder, to encode the chemical, structural, and bond angular information of a crystal graph $\mathcal{G}$. We derive additional line graph $\mathcal{L}(\mathcal{G})$ from the crystal graph $\mathcal{G}$ to describe the angles between the edges in $\mathcal{G}$, where nodes and edges in line graph $\mathcal{L}(\mathcal{G})$ correspond to inter-atomic bonds and bond angles. We denote $h^l_i$, $e^l_{i,j}$ and $t^l_{i,j,k}$ as \textit{l}-th layer representation for i-th atom, $\{i,j\}$-th bond, and $\{i,j,k\}$-th angle (triplet) respectively. Graph encoder alternates edge-gated graph convolution layers between $\mathcal{L}(\mathcal{G})$ and $\mathcal{G}$ to propagate bond angular information through inter-atomic bond representation to atom embedding and vice versa. Specifically, at the $(l)$-th layer, given the line graph $\mathcal{L}(\mathcal{G})$, we apply Gated Graph ConvNet (GatedGCN) ~\cite{dwivedi2020benchmarkgnns} to update triplet representation and generate bond messages m as follows :
\begin{equation}
    \label{eq:linegraph}
    \begin{split}
    & {t}^{l+1}_{i,j,k} = {t}^{l}_{i,j,k} + \gamma \biggl(BN\biggl( A^{l}_{lg}e_{i,j}^{l} + B^{l}_{lg}e_{j,k}^{l} + C^{l}_{lg}{t}^{l}_{i,j,k}\biggr)\biggr) \\ 
    & \hat{t}^{l+1}_{i,j,k} = \frac{\sigma (t^{l+1}_{i,j,k})}{\sum_{\substack{(j,m) \in N_{i,j}}} \sigma (t^{l+1}_{i,j,m}) + \epsilon }\\
    & m_{i,j}^{l} = e_{i,j}^{l} + \gamma \biggl(BN\biggl((W^{l}_{lg}e_{i,j}^{l} +\sum_{\substack{(j,k) \\ \in N_{i,j}}}\hat{t}^{l+1}_{i,j,k} \odot V^{l}_{lg}e_{j,k}^{l}\biggr)\biggr)
    \end{split}
\end{equation}
Further, we apply another GatedGCN on the crystal graph $\mathcal{G}$ and update bond and atom features as follows : 
\begin{equation}
    \label{eq:crystalgraph}
    \begin{split}
    & {e}^{l+1}_{i,j} = {e}^{l}_{i,j} + \gamma \biggl(BN\biggl( A^{l}_{g}h_{i}^{l} + B^{l}_{g}h_{j}^{l} + C^{l}_{g}{m}^{l}_{i,j}\biggr)\biggr) \\
    & \hat{e}^{l+1}_{i,j} = \frac{\sigma (e^{l+1}_{i,j})}{\sum_{\substack{k \in N_{i}}} \sigma (e^{l+1}_{i,k}) + \epsilon }\\
    & h^{l+1}_{i} = h^{l}_{i} + \gamma \biggl(BN\biggl((W^{l}_{g}h^{l}_{i} +\sum_{\substack{j \in N_{i}}}\hat{e}^{l+1}_{i,j} \odot V^{l}_{g}h_{j}^{l}\biggr)\biggr)
    \end{split}
\end{equation}
where $\sigma$ is the sigmoid function, $\epsilon$ is a small fixed constant for numerical stability, $\odot$ is the Hadamard product, BN is batch normalization and $\gamma$ is the activation function where we use Sigmoid Linear Unit (SiLU). $A^{l}_{lg},B^{l}_{lg},C^{l}_{lg},V^{l}_{lg},W^{l}_{lg}$ are learnable parameters of GatedGCN applied on $\mathcal{L}(\mathcal{G})$ and $A^{l}_{g},B^{l}_{g},C^{l}_{g},V^{l}_{g},W^{l}_{g}$ are learnable parameters of GatedGCN applied on $\mathcal{G}$. We apply $L$ such layers of aggregation and update in Graph Encoder and return the final set of node embeddings $\mathcal{H} = \{h_1,..., h_{|\mathcal{V}|}\}$, where $h_i:=h_i^L \in R^{d}$ represents the final embedding of node $i$. We subsequently use a symmetric aggregation function (AvgPool) to generate graph-level representation $\mathcal{Z}_{\mathcal{G}} \in R^{d'}$ (we set $d'$ as 256) of the crystal material M.
\begin{equation}
    \label{eq:aggregation}
    \begin{split}
    & \mathcal{Z}_{\mathcal{G}} = \sum_{i=1}^{|\mathcal{V}|}  h_{i}^{L}
    \end{split}
\end{equation}
\begin{table*}
  \centering
    \setlength{\tabcolsep}{3 pt}
      \begin{tabular}{c c | c c c c c c c c c}
        \toprule
        Property &  Unit & CIFID & CGCNN & SchNet & MEGNET & GATGNN & ALIGNN  & Matformer & \our{} \\
        \midrule
        Formation Energy  & eV/atom & 0.140   & 0.063  & 0.045  & 0.047 & 0.047  & 0.033           & 0.033\textbf{*} &\textbf{0.028}\\
        Bandgap(OPT)      & eV      & 0.301   & 0.200  & 0.192  & 0.145 & 0.170  & 0.142           & 0.137\textbf{*}  &\textbf{0.128}\\
        Bandgap(MBJ)      & eV      & 0.532   & 0.413  & 0.433  & 0.344 & 0.513  & 0.310           & 0.302\textbf{*}  & \textbf{0.278}\\
        Total Energy      & eV/atom & 0.244   & 0.078  & 0.047  & 0.058 & 0.056  & 0.037           & 0.035\textbf{*} & \textbf{0.034}\\
        Bulk Moduli(Kv)   & GPa     & 14.12   & 14.47  & 14.33  & 15.11 & 14.32  & 10.40\textbf{*} & 11.21 & \textbf{9.625}\\
        Shear Moduli(Gv)  & GPa     & 11.98   & 11.75  & 10.67  & 13.09 & 12.48  & 9.481\textbf{*}  & 10.76 & \textbf{8.471}\\
         \bottomrule
      \end{tabular} 
   \caption{Summary of the prediction performance (MAE) of \our{} and different state-of-the-art models for different properties in JARVIS-DFT Dataset. The best performance is highlighted in bold and the second-best results are highlighted with \textbf{*}.}
  \label{tbl-jarvis}
\end{table*}
\subsubsection{Text Encoder: }As a text encoder, we adopt a pre-trained MatSciBERT model, which is a domain-specific language model for materials science, followed by a projection layer. MatSciBERT is effectively a pre-trained SciBERT model on a scientific text corpus of 3.17B words, which is further trained on a huge text corpus of materials science containing around 285 M words, using domain adaptive pretraining objective proposed by \cite{gururangan2020don}. We feed textual description of material $\mathcal{T}$ and extract embedding of [CLS] token $\mathcal{Z}_{CLS} \in R^{768}$ as a representation of the whole text. Further. we pass $\mathcal{Z}_{CLS}$ through a projection layer (two-layer neural network) to generate the textual embedding for the material $\mathcal{Z}_{\mathcal{T}} \in R^{d}$
\begin{equation}
    \label{eq:projection_head}
    \begin{split}
    & \mathcal{Z}_{\mathcal{T}} = W_2(g(W_1\mathcal{Z}_{CLS}))
    \end{split}
\end{equation}
We use standard non-linear function ReLU(·) as g(·), $W_1 \in R^{768 \times 128}$ and $W_2 \in R^{128 \times d}$ are parameter matrix that project $\mathcal{Z}_{CLS}$ to embedding space  $R^{d}$. In our experiment, we set d as 64.
\subsubsection{Joint Embedding Model: } The graph encoder encodes local structural and chemical semantics around atoms in a unit cell of the material, whereas the text encoder captures global periodic knowledge from the textual description. Further, in the joint embedding model, we fuse both the representations $(\mathcal{Z}_{\mathcal{G}},\mathcal{Z}_{\mathcal{T}})$ together into a single multi-modal representation $\mathcal{Z}:=  (\mathcal{Z}_{\mathcal{G}} \oplus \mathcal{Z}_{\mathcal{T}}) \in R^{(d'+d)}$, which can now capture both local and global structural semantics of the material. We tried different ways to fuse both the embeddings like sum, average, concatenation $(\oplus)$ and found concatenation performs best.\\
Further, we pass this multi-modal representation $\mathcal{Z}$ through a multi-layer perceptron which predicts the
value of the properties. We train \our{} end to end to optimize the following mean square error(MSE) loss :
\begin{equation}
\label{eq:mse_loss}
                \mathcal{L}_{MSE}=  \lVert {\hat{\mathcal{Y}}} - {\mathcal{Y}} \rVert^2
\end{equation}
where $\hat{\mathcal{Y}}$ and $\mathcal{Y}$ are predicted and true property values respectively. Note, while training \our{} we freeze the weights of MatSciBERT and don't tune it further. Fine-tuning MatSciBERT with \our{} training for a specific property will add more computational overhead as it will increase the number of parameters significantly. This provides scope for further investigation and we keep it as future work.

\begin{table*}
  \centering
    \setlength{\tabcolsep}{4.5 pt}
      \begin{tabular}{c c | c c c c c c c }
        \toprule
        Property &  Unit  & CGCNN & SchNet & MEGNET & GATGNN & ALIGNN & Matformer &\our{} \\
        \midrule
        Formation Energy  & eV/atom  &  0.031 & 0.033  & 0.030 & 0.033 & 0.022 & 0.021\textbf{*} & \textbf{0.020}\\
        Bandgap           & eV       &  0.292 & 0.345  & 0.307 & 0.280 & 0.218 & 0.211\textbf{*} & \textbf{0.197}\\
        Bulk Moduli(Kv)   & log(GPa) &  0.047 & 0.066  & 0.060 & 0.045 & 0.051 & 0.043\textbf{*} & \textbf{0.038}\\
        Shear Moduli(Gv)  & log(GPa) &  0.077 & 0.099  & 0.099 & 0.075 & 0.078 & 0.073\textbf{*} & \textbf{0.062}\\
         \bottomrule
      \end{tabular} 
   \caption{Summary of the prediction performance (MAE) of \our{} and different state-of-the-art models for different properties in The Materials Project dataset. The best performance is highlighted in bold and the second-best results are highlighted with \textbf{*}.}
  \label{tbl-mp}
\end{table*}
\begin{table}
  \centering
  \small
    \setlength{\tabcolsep}{3 pt}
      \begin{tabular}{c c | c c c}
        \toprule
        Property & Train-set &CGCNN & ALIGNN & \our{} \\
         & Size & &  &  \\
        \midrule
        Bandgap(MBJ) & 3634 & 0.522  & 0.483 & \textbf{0.456} \\
        Bulk Moduli  & 3936 &  14.98 & 14.13 & \textbf{13.04} \\
        Shear Moduli & 3936 &  13.07 & 12.61 & \textbf{11.57} \\
        \bottomrule
      \end{tabular} 
   \caption{: MAE values of CGCNN, ALIGNN and \our{} for three different properties in the JARVIS-DFT dataset with 20\% training instances. The best performance is highlighted in bold.}
  \label{tbl-limited-data}
\end{table}
\section{Experimental Results}
In this section, we begin by describing the experimental setup which includes the benchmark datasets used for evaluation, alternate baseline approaches, and implementation details.  Then we evaluate the performance of \our{} in comparison with different SOTA property predictors on the downstream property prediction tasks using two popular material benchmark datasets. Next, we present the empirical evaluation results of our proposed framework in limited training data settings. Further, we conduct some ablation studies to demonstrate the expressiveness and robustness of textual representation and the importance of global and local knowledge encoded in textual embedding. Finally, we perform a qualitative analysis of the attention layer of MatSciBert to visualize attention in different tokens in the material description.
\subsection{Experimental Setup}
To evaluate the effectiveness of \our{}, we conduct experiments on two benchmark material datasets, Materials Project ~\cite{MP} (MP 2018.6.1) and JARVIS-DFT ~\cite{choudhary2020joint} (2021.8.1), which comprises some important physical properties obtained with high-throughput DFT calculations. 
MP 2018.6.1 consists of 69,239 materials  whereas JARVIS-DFT consists of 55,722 materials. We curated textual datasets for both datasets using robocrystallographer with a textual description of each material as described in subsection \ref{textual}. 
We choose seven state of the art algorithms for crystal property prediction CIFID ~\cite{choudhary2018machine}, CGCNN ~\cite{xie2018crystal}, SchNet ~\cite{schutt2017schnet}, MEGNET ~\cite{chen2019graph}, GATGNN ~\cite{louis2020graph}, ALIGNN ~\cite{choudhary2021atomistic} and Matformer ~\cite{yan2022periodic}. To avoid any deterioration of the performance of the baseline algorithms due to insufficient hyperparameter tuning, we report the property prediction results from the respective papers of the baseline models.\\
We use four convolution layers of the graph encoder module and pre-trained MatSciBERT followed by a two-layer neural network (projection layer) as the text encoder module in \our{}. We train it for 1000 epochs using AdamW  \cite{loshchilov2017decoupled} optimizer with normalized weight decay of $10^{-5}$ and keep the batch size as 64. We schedule the learning rate according to the one-cycle policy ~\cite{smith2018disciplined} with a maximum learning rate of 0.001. We keep embedding dimensions of  the graph and text encoder as 64 and 256 respectively. We perform the experiments in shared servers having Intel E5-2620v4 processors which contain 16 cores/thread and four GTX 1080Ti 11GB GPUs each.
\subsection{Downstream Task Evaluation}
\subsubsection{JARVIS-DFT Dataset}
To evaluate \our{}, we first conduct experiments on the JARVIS-DFT dataset, which is a widely used large-scale material benchmark containing 55,722 crystals. Following previous state-of-the-art works, we choose six crystal properties including formation energy, bandgap (OPT), bandgap (MBJ), total energy, bulk moduli, and shear moduli for the downstream property prediction task. We use 80\%,10\%, and 10\% train, validation, and test split for all the properties as used by ALIGNN. We report mean absolute error (MAE) of the predicted and actual value of a particular property for test data in table \ref{tbl-jarvis} to compare the performance of \our{} and different participating methods.
We observe that \our{} outperforms every baseline model across all the properties with a significant margin. In specific, we observe 13.84\%, 2.85\%, 6.56\%, 7.33\%, 7.40\%, and 10.65\%  improvements compared to the competing second-best baseline model for formation energy, total energy, bandgap (OPT), bandgap (MBJ), bulk moduli, and shear moduli respectively, which shows the effectiveness of multimodal representation capturing both local chemical semantics and global periodic structural knowledge towards crystal property prediction.
\subsubsection{Materials Project (MP) Dataset}
We further use another benchmark material dataset, Materials Project-2018.6.1, comprises 69,239 materials. Here we evaluate \our{} with all state-of-the-art models using four crystal properties namely formation energy, bandgap, bulk moduli, and shear moduli. For formation energy and bandgap, we use 60000, 5000, and 4239 crystals as train, validation, and test split as used by ALIGNN, whereas use 4664, 393, and 393 crystals as train, validation, and test split for bulk and shear moduli as used by GATGNN. We report the mean absolute error (MAE) of the predicted and actual property value for test data in table \ref{tbl-mp} to compare the performance of \our{} with different participating methods. Note, to maintain consistency with the results reported by baseline works, we report (GPa) values of bulk and shear moduli in table \ref{tbl-jarvis}, whereas log(GPa) values in table \ref{tbl-mp}. We observe that \our{} outperforms every baseline model across all the properties with a significant margin. In specific, we observe 4.76\%, 6.63\%, 6.9\%, and 15.06\%  improvements compared to the competing second-best baseline model for formation energy, bandgap, bulk moduli, and shear moduli respectively. Overall, the superior performances show the effectiveness of multi-modal representation in \our{}.

\subsubsection{Resuts on Limited Training Data}
\our{} performs well in limited data settings as well. With 4664 training samples only \our{} achieves 6.9\% and 15.06\% improvements for bulk moduli and shear moduli respectively in the Materials Project dataset. Further, in JARVIS-DFT Dataset, we conduct an additional set of experiments for three different properties including bandgap (MBJ), bulk moduli, and shear moduli, where we have limited labeled data.  More specifically, we take 20-10-10\% training-validation-test data split and evaluate the performance of CGCNN, ALIGNN, and \our{} in the table \ref{tbl-limited-data}. We observe, \our{} achieves improvement for all three properties compared to CGCNN and ALIGNN. Overall, these superior performances indicate the robustness of our model and its adaptive ability to tasks of various data scales.
\begin{table}
  \centering
  \small
    \setlength{\tabcolsep}{6 pt}
      \begin{tabular}{c  c c c}
        \toprule
        Property & CrysTextNet & CGCNN & ALIGNN \\
        \midrule
        Formation Energy  & 0.447  &  0.063 & 0.033 \\
        Total Energy      & 0.352  &  0.078 & 0.037 \\
        \midrule
        Bandgap(OPT)      & 0.595  &  0.201 & 0.142 \\
        Bandgap(MBJ)      & 0.849  &  0.411 & 0.311 \\
        \midrule
        Bulk Moduli(Kv)   & 21.98  &  14.47 & 10.42 \\
        Shear Moduli(Gv)  & 14.76 &   11.75 & 9.483 \\
         \bottomrule
      \end{tabular} 
      \caption{Summary of experiments for the ablation study on the effectiveness of Textual Representation.}
  \label{tbl-onlytext}
\end{table}
\begin{table*}
  \centering
  \small
    \setlength{\tabcolsep}{11 pt}
      \begin{tabular}{c | c >{\columncolor{green!20}}c | c >{\columncolor{green!20}}c | c >{\columncolor{green!20}}c}
        \toprule
        Property & CGCNN & CrysMMNet & MEGNET & CrysMMNet & GATGNN & CrysMMNet\\
         &  & (CGCNN) &  & (MEGNET) &  & (GATGNN)\\
        \midrule
        Formation Energy  & 0.063  &  0.046 & 0.076 & 0.060 & 0.077 & 0.064 \\
        Bandgap(OPT)      & 0.200  &  0.163 & 0.184 & 0.165 & 0.169 & 0.157 \\
        Bandgap(MBJ)      & 0.413   &  0.339 & 0.369 & 0.339	& 0.343 & 0.331 \\
        Total Energy      & 0.078  &  0.059 & 0.058 &  0.057     & 0.056 &  0.053     \\
        Bulk Moduli(Kv)   & 14.47  &  12.98 & 15.11 & 13.29 & 14.32 & 13.73 \\
        Shear Moduli(Gv)  & 11.75 &   10.71 & 13.09 & 11.86 & 12.48 & 12.04 \\
         \bottomrule
      \end{tabular} 
   \caption{Summary of the prediction performance (MAE) of different state-of-the-art GNN models with textual representation for six different properties in The JARVIS-DFT Dataset. Model M is the SOTA baseline model and \our{}(M) is a variant where we replace graph encoder with M.}
  \label{tbl-all-gnn}
\end{table*}
\subsection{Ablation Study} In this subsection, We demonstrate the robustness of multimodal representation on different GNN architecture choices and the influence of textual modality on \our{} performance, by designing the following set of ablation studies:
\begin{enumerate}
    \item Is only textual information sufficient to infer better property prediction accuracy?
    \item How robust is the multimodal representation on different GNN architecture choices for graph encoder?
    \item What are the influences of global structural and local compositional knowledge from the textual datasets on property prediction performance?
\end{enumerate}
In the following subsections, we will thoroughly discuss these.
\subsubsection{Expressiveness of Textual Representation}
First, we are interested to understand whether textual representations are alone expressive enough, to encode atomic chemical and periodic structural semantics from the curated textual data and predict different properties precisely. We conduct an ablation experiment, where we consider only the text embeddings $\mathcal{Z}_{\mathcal{T}}$ of \our{} (output of projection layer) and pass it alone through a multi-layer perceptron to predict the property value. We denote this model as CrysTextNet and compare it with state-of-the-art graph-based models on different properties of the JARVIS-DFT dataset. We report the MAE for test data in table \ref{tbl-onlytext}. We observe, for all the properties, test MAE is higher for the CrysTextNet model compared to state-of-the-art graph-based models like CGCNN and ALIGNN. In specific, for mechanical properties like  bulk modulus and shear modulus, CrysTextNet works better (closer test MAE with competing GNN baselines) than properties like formation energy, band-gap, and total energy. This is because properties, such as formation energy, band gap, and total energy rely on microscopic chemical information which textual representation fails to encode. Instead, graph encodings include node features ${\bf x}_u$ ($u \in \mathcal{V}$) that are high-dimensional vectors with meaningful chemical quantities like electronegativity, group number, covalent radius, number of valence electrons, first ionization energy, etc. Furthermore, through message passing and aggregation in the graph convolution layer, GNN models capture many-body interactions among atoms in the material. On the other side, mechanical properties like bulk modulus and shear modulus are more dependent on structural information like lattice structure and symmetry of the material, which textual representations are able to capture.\\
Overall, though textual representations can capture many useful local and global information about the materials, unlike graph structural models, they are not alone expressive enough to capture atomic chemical features and the structural connectivity between different atoms in the materials.  Message passing and neighborhood aggregation between atoms through the GNN model are still very fundamental in learning the structure-property relationship of the materials.
\subsubsection{Robustness of Textual Representation}
Further, we investigate the robustness of textual representations on different crystal GNN encoders. We conduct an ablation study where we replace graph encode of \our{} with popular crystal GNN variants, e.g, CGCNN, MEGNET, GATGNN  and evaluate the performance. We set up the experiments with six properties of the JARVIS dataset and report the MAE values in table \ref{tbl-all-gnn} for the baseline GNN models and different variants of \our{} with different GNN architectural choices as graph encoder. We observe all these variants outperform corresponding vanilla GNN models with a good margin for all the properties, which shows textual representations are rich enough to encode global structural knowledge which aids the property prediction accuracy of any state-of-the-art GNN models. 
\subsubsection{Importance of Local and Global Knowledge}
Finally, we are curious to understand the importance of local (atom/bond) compositional information and global material structural knowledge encoded through textual representation in \our{}. Specifically, we conduct an ablation study, where we train \our{} in two additional setups along with the conventional (Global+Local) setup for \our{}. (a) Only Global: In this scenario, we take only global knowledge about the periodic structure of the materials as textual data to train \our{}. (b) Only Local: In this scenario, we take only local compositional information about atoms and inter-atomic bonds as textual data to train \our{}. We use six properties of JARVIS-DFT dataset for the experiment and report MAE in Table \ref{tbl-text-ablation}. We observe performance gain across all the properties using both global and local information combined as textual knowledge, compared to only global or local knowledge separately. 
\begin{table}
  \centering
  \small
    \setlength{\tabcolsep}{3 pt}
      \begin{tabular}{c c c c}
        \toprule
        Property & Global+Local & Only Global & Only Local \\
        \midrule
        Formation Energy  & \textbf{0.028}  &  0.039 & 0.039 \\
        Total Energy      & \textbf{0.034}  &  0.042 & 0.046 \\
        \midrule
        Bandgap(OPT)      & \textbf{0.114}  &  0.191 & 0.147 \\
        Bandgap(MBJ)      & \textbf{0.209}  &  0.216 & 0.218 \\
        \midrule
        Bulk Moduli(Kv)   & \textbf{6.860}  &  6.910 & 6.870 \\
        Shear Moduli(Gv)  & \textbf{6.440} &   6.730 & 6.880 \\
         \bottomrule
      \end{tabular} 
   \caption{Summary of experiments for the ablation study on the importance of Local and Global Material Knowledge.}
  \label{tbl-text-ablation}
\end{table}
\subsection{Qualitative Analysis of Attention Layers}Finally, to visualize and understand attentions in different tokens in the material description, we perform a qualitative analysis of the attention layer in MatSciBert.  We utilized the standard BertViz tool ~\cite{vig-2019-multiscale} \footnote{https://github.com/jessevig/bertviz} to analyze and visualize the attention scores in the MatSciBert Model. We present a case study of  the textual data of $FeH_8(ClO_2)_2$ in \textit{\textbf{Figure 3 \& 4 in Section A of Appendix}}, where we have examined the attention score of the [CLS] token at the 5th layer of MatSciBert. \\
We observe MatSciBert allocates higher attention scores to tokens that defines global features of the crystal, such as   \textit{`Formula', `Mineral', `Crystal System', `Space Group Number', and `Dimensionality'}. Further, we investigate attention weights for local information corresponding to Fe, H, and O atoms. MatScibert provides more attention score to tokens related to \textit{bond types (octahedral geometry, equivalent bond, distorted water-like geometry, etc)} and \textit{bond lengths (2.08 Å, 2.10 Å, and 2.53 Å bond length)}. It is evident from these observations that MatSciBert is attending the important tokens related to global and local material information, to generate more expressive multimodal representation.
\section{Conclusions}
In this work, we address the limitation of state-of-the-art GNN models for crystal property prediction  to capture global periodic structural information and leverage textual modalities beside graph structures to resolve the issue. To this end, we curate textual datasets of two popular benchmark databases containing textual descriptions of each material containing both local compositional and global structural information of a material. Further,  we propose a simple yet effective multi-modal framework, \our{}, for crystalline materials, which fuse both graph structural and textual representation together to generate a more enriched and robust multimodal representation for materials, which subsequently improves property prediction accuracy. Extensive experiments show \our{} outperforms all the popular state-of-the-art baselines across ten diverse sets of properties on two popular datasets. Further, we conduct ablation studies to demonstrate the expressiveness and robustness of textual representation on different crystal GNN encoders and show performance gain across all the properties using both global and local information combined as textual knowledge, compared to only global or local knowledge separately. Finally, we visualize attention weights between [CLS] token and other tokens in the material's description to understand
the important tokens. that the text encoder is attending to generate more expressive multimodal representation
\section{Acknowledgments}
This work was partially funded by Indo Korea Science and Technology Center, Bangalore, India, under the project name {\em Transfer learning and Weak Supervision for Accurate and Interpretable Prediction of Properties of Materials from their Crystal Graph Representation} and the Federal Ministry of Education and Research (BMBF), Germany under the project ``LeibnizKILabor'' with grant No. 01DD20003.  We thank the Ministry of Education, Govt of India, for supporting Kishalay with Prime Minister Research Fellowship during his Ph.D. tenure.
\bibliography{das_605}

\begin{thebibliography}{41}
\providecommand{\natexlab}[1]{#1}
\providecommand{\url}[1]{\texttt{#1}}
\expandafter\ifx\csname urlstyle\endcsname\relax
  \providecommand{\doi}[1]{doi: #1}\else
  \providecommand{\doi}{doi: \begingroup \urlstyle{rm}\Url}\fi

\bibitem[Bandyopadhyay et~al.(2020)Bandyopadhyay, Das, and
  Murty]{bandyopadhyay2020hypergraph}
Sambaran Bandyopadhyay, Kishalay Das, and M~Narasimha Murty.
\newblock Hypergraph attention isomorphism network by learning line graph
  expansion.
\newblock In \emph{2020 IEEE International Conference on Big Data (Big Data)},
  pages 669--678. IEEE, 2020.

\bibitem[Berg et~al.(2017)Berg, Kipf, and Welling]{berg2017graph}
Rianne van~den Berg, Thomas~N Kipf, and Max Welling.
\newblock Graph convolutional matrix completion.
\newblock \emph{arXiv preprint arXiv:1706.02263}, 2017.

\bibitem[Chen et~al.(2019)Chen, Ye, Zuo, Zheng, and Ong]{chen2019graph}
Chi Chen, Weike Ye, Yunxing Zuo, Chen Zheng, and Shyue~Ping Ong.
\newblock Graph networks as a universal machine learning framework for
  molecules and crystals.
\newblock \emph{Chem. Mater.}, 31\penalty0 (9):\penalty0 3564--3572, 2019.

\bibitem[Chen et~al.(2018)Chen, Ma, and Xiao]{chen2018fastgcn}
Jie Chen, Tengfei Ma, and Cao Xiao.
\newblock Fastgcn: fast learning with graph convolutional networks via
  importance sampling.
\newblock \emph{arXiv preprint arXiv:1801.10247}, 2018.

\bibitem[Choudhary and DeCost(2021)]{choudhary2021atomistic}
Kamal Choudhary and Brian DeCost.
\newblock Atomistic line graph neural network for improved materials property
  predictions.
\newblock \emph{npj Computational Materials}, 7\penalty0 (1):\penalty0 1--8,
  2021.

\bibitem[Choudhary et~al.(2018)Choudhary, DeCost, and
  Tavazza]{choudhary2018machine}
Kamal Choudhary, Brian DeCost, and Francesca Tavazza.
\newblock Machine learning with force-field-inspired descriptors for materials:
  Fast screening and mapping energy landscape.
\newblock \emph{Physical review materials}, 2\penalty0 (8):\penalty0 083801,
  2018.

\bibitem[Choudhary et~al.(2020)Choudhary, Garrity, Reid, DeCost, Biacchi,
  Hight~Walker, Trautt, Hattrick-Simpers, Kusne, Centrone,
  et~al.]{choudhary2020joint}
Kamal Choudhary, Kevin~F Garrity, Andrew~CE Reid, Brian DeCost, Adam~J Biacchi,
  Angela~R Hight~Walker, Zachary Trautt, Jason Hattrick-Simpers, A~Gilad Kusne,
  Andrea Centrone, et~al.
\newblock The joint automated repository for various integrated simulations
  (jarvis) for data-driven materials design.
\newblock \emph{npj computational materials}, 6\penalty0 (1):\penalty0 173,
  2020.

\bibitem[Dai et~al.(2018)Dai, Kozareva, Dai, Smola, and Song]{dai2018learning}
Hanjun Dai, Zornitsa Kozareva, Bo~Dai, Alex Smola, and Le~Song.
\newblock Learning steady-states of iterative algorithms over graphs.
\newblock In \emph{International conference on machine learning}, pages
  1106--1114. PMLR, 2018.

\bibitem[Das et~al.(2022)Das, Samanta, Goyal, Lee, Bhattacharjee, and
  Ganguly]{das2022crysxpp}
Kishalay Das, Bidisha Samanta, Pawan Goyal, Seung-Cheol Lee, Satadeep
  Bhattacharjee, and Niloy Ganguly.
\newblock Crysxpp: An explainable property predictor for crystalline materials.
\newblock \emph{npj Computational Materials}, 8\penalty0 (1):\penalty0 1--11,
  2022.

\bibitem[Das et~al.(2023)Das, Samanta, Goyal, Lee, Bhattacharjee, and
  Ganguly]{das2023crysgnn}
Kishalay Das, Bidisha Samanta, Pawan Goyal, Seung-Cheol Lee, Satadeep
  Bhattacharjee, and Niloy Ganguly.
\newblock Crysgnn: Distilling pre-trained knowledge to enhance property
  prediction for crystalline materials.
\newblock \emph{Proceedings of the AAAI Conference on Artificial Intelligence},
  2023.

\bibitem[De~Jong et~al.(2016)De~Jong, Chen, Notestine, Persson, Ceder, Jain,
  Asta, and Gamst]{de2016statistical}
Maarten De~Jong, Wei Chen, Randy Notestine, Kristin Persson, Gerbrand Ceder,
  Anubhav Jain, Mark Asta, and Anthony Gamst.
\newblock A statistical learning framework for materials science: application
  to elastic moduli of k-nary inorganic polycrystalline compounds.
\newblock \emph{Scientific reports}, 6\penalty0 (1):\penalty0 1--11, 2016.

\bibitem[Duvenaud et~al.(2015)Duvenaud, Maclaurin, Iparraguirre, Bombarell,
  Hirzel, Aspuru-Guzik, and Adams]{duvenaud2015convolutional}
David~K Duvenaud, Dougal Maclaurin, Jorge Iparraguirre, Rafael Bombarell,
  Timothy Hirzel, Al{\'a}n Aspuru-Guzik, and Ryan~P Adams.
\newblock Convolutional networks on graphs for learning molecular fingerprints.
\newblock \emph{Advances in neural information processing systems}, 28, 2015.

\bibitem[Dwivedi et~al.(2020)Dwivedi, Joshi, Luu, Laurent, Bengio, and
  Bresson]{dwivedi2020benchmarkgnns}
Vijay~Prakash Dwivedi, Chaitanya~K Joshi, Anh~Tuan Luu, Thomas Laurent, Yoshua
  Bengio, and Xavier Bresson.
\newblock Benchmarking graph neural networks.
\newblock \emph{arXiv preprint arXiv:2003.00982}, 2020.

\bibitem[Ganose and Jain(2019)]{ganose2019robocrystallographer}
Alex~M Ganose and Anubhav Jain.
\newblock Robocrystallographer: automated crystal structure text descriptions
  and analysis.
\newblock \emph{MRS Communications}, 9\penalty0 (3):\penalty0 874--881, 2019.

\bibitem[Gilmer et~al.(2017)Gilmer, Schoenholz, Riley, Vinyals, and
  Dahl]{gilmer2017neural}
Justin Gilmer, Samuel~S Schoenholz, Patrick~F Riley, Oriol Vinyals, and
  George~E Dahl.
\newblock Neural message passing for quantum chemistry.
\newblock In \emph{International conference on machine learning}, pages
  1263--1272. PMLR, 2017.

\bibitem[Gururangan et~al.(2020)Gururangan, Marasovi{\'c}, Swayamdipta, Lo,
  Beltagy, Downey, and Smith]{gururangan2020don}
Suchin Gururangan, Ana Marasovi{\'c}, Swabha Swayamdipta, Kyle Lo, Iz~Beltagy,
  Doug Downey, and Noah~A Smith.
\newblock Don't stop pretraining: Adapt language models to domains and tasks.
\newblock \emph{arXiv preprint arXiv:2004.10964}, 2020.

\bibitem[Hamilton et~al.(2017)Hamilton, Ying, and
  Leskovec]{hamilton2017inductive}
Will Hamilton, Zhitao Ying, and Jure Leskovec.
\newblock Inductive representation learning on large graphs.
\newblock \emph{Advances in neural information processing systems}, 30, 2017.

\bibitem[Hsu et~al.(2021)Hsu, Pham, Keilbart, Weitzner, Chapman, Xiao, Qiu,
  Chen, and Wood]{hsu2021efficient}
Tim Hsu, Tuan~Anh Pham, Nathan Keilbart, Stephen Weitzner, James Chapman,
  Penghao Xiao, S~Roger Qiu, Xiao Chen, and Brandon~C Wood.
\newblock Efficient, interpretable graph neural network representation for
  angle-dependent properties and its application to optical spectroscopy.
\newblock \emph{arXiv preprint arXiv:2109.11576}, 2021.

\bibitem[Im et~al.(2019)Im, Lee, Ko, Kim, Hyon, and Chang]{im2019identifying}
Jino Im, Seongwon Lee, Tae-Wook Ko, Hyun~Woo Kim, YunKyong Hyon, and Hyunju
  Chang.
\newblock Identifying pb-free perovskites for solar cells by machine learning.
\newblock \emph{npj Computational Materials}, 5\penalty0 (1):\penalty0 1--8,
  2019.

\bibitem[Isayev et~al.(2017)Isayev, Oses, Toher, Gossett, Curtarolo, and
  Tropsha]{isayev2017universal}
Olexandr Isayev, Corey Oses, Cormac Toher, Eric Gossett, Stefano Curtarolo, and
  Alexander Tropsha.
\newblock Universal fragment descriptors for predicting properties of inorganic
  crystals.
\newblock \emph{Nature communications}, 8\penalty0 (1):\penalty0 1--12, 2017.

\bibitem[Jain et~al.(2013)Jain, Ong, Hautier, Chen, Richards, Dacek, Cholia,
  Gunter, Skinner, Ceder, et~al.]{MP}
Anubhav Jain, Shyue~Ping Ong, Geoffroy Hautier, Wei Chen, William~Davidson
  Richards, Stephen Dacek, Shreyas Cholia, Dan Gunter, David Skinner, Gerbrand
  Ceder, et~al.
\newblock Commentary: The materials project: A materials genome approach to
  accelerating materials innovation.
\newblock \emph{APL materials}, 1\penalty0 (1):\penalty0 011002, 2013.

\bibitem[Jha et~al.(2019)Jha, Choudhary, Tavazza, Liao, Choudhary, Campbell,
  and Agrawal]{jha2019enhancing}
Dipendra Jha, Kamal Choudhary, Francesca Tavazza, Wei-keng Liao, Alok
  Choudhary, Carelyn Campbell, and Ankit Agrawal.
\newblock Enhancing materials property prediction by leveraging computational
  and experimental data using deep transfer learning.
\newblock \emph{Nature communications}, 10\penalty0 (1):\penalty0 1--12, 2019.

\bibitem[Larsen et~al.(2019)Larsen, Pandey, Strange, and Jacobsen]{Larsen}
Peter~Mahler Larsen, Mohnish Pandey, Mikkel Strange, and Karsten~Wedel
  Jacobsen.
\newblock Definition of a scoring parameter to identify low-dimensional
  materials components.
\newblock \emph{Physical Review Materials}, 3\penalty0 (3):\penalty0 034003,
  2019.

\bibitem[Lee et~al.(2016)Lee, Seko, Shitara, Nakayama, and
  Tanaka]{lee2016prediction}
Joohwi Lee, Atsuto Seko, Kazuki Shitara, Keita Nakayama, and Isao Tanaka.
\newblock Prediction model of band gap for inorganic compounds by combination
  of density functional theory calculations and machine learning techniques.
\newblock \emph{Physical Review B}, 93\penalty0 (11):\penalty0 115104, 2016.

\bibitem[Loshchilov and Hutter(2017)]{loshchilov2017decoupled}
Ilya Loshchilov and Frank Hutter.
\newblock Decoupled weight decay regularization.
\newblock \emph{arXiv preprint arXiv:1711.05101}, 2017.

\bibitem[Louis et~al.(2020)Louis, Zhao, Nasiri, Wang, Song, Liu, and
  Hu]{louis2020graph}
Steph-Yves Louis, Yong Zhao, Alireza Nasiri, Xiran Wang, Yuqi Song, Fei Liu,
  and Jianjun Hu.
\newblock Graph convolutional neural networks with global attention for
  improved materials property prediction.
\newblock \emph{Physical Chemistry Chemical Physics}, 22\penalty0
  (32):\penalty0 18141--18148, 2020.

\bibitem[Lu et~al.(2018)Lu, Zhou, Ouyang, Guo, Li, and Wang]{lu2018accelerated}
Shuaihua Lu, Qionghua Zhou, Yixin Ouyang, Yilv Guo, Qiang Li, and Jinlan Wang.
\newblock Accelerated discovery of stable lead-free hybrid organic-inorganic
  perovskites via machine learning.
\newblock \emph{Nature communications}, 9\penalty0 (1):\penalty0 1--8, 2018.

\bibitem[Park and Wolverton(2020)]{Wolverton2020}
Cheol~Woo Park and Chris Wolverton.
\newblock Developing an improved crystal graph convolutional neural network
  framework for accelerated materials discovery.
\newblock \emph{Physical Review Materials}, 4\penalty0 (6), Jun 2020.
\newblock ISSN 2475-9953.
\newblock \doi{10.1103/physrevmaterials.4.063801}.
\newblock URL \url{http://dx.doi.org/10.1103/PhysRevMaterials.4.063801}.

\bibitem[Pilania et~al.(2015)Pilania, Gubernatis, and
  Lookman]{pilania2015structure}
Ghanshyam Pilania, James~E Gubernatis, and TJPRB Lookman.
\newblock Structure classification and melting temperature prediction in octet
  ab solids via machine learning.
\newblock \emph{Physical Review B}, 91\penalty0 (21):\penalty0 214302, 2015.

\bibitem[Schmidt et~al.(2021)Schmidt, Pettersson, Verdozzi, Botti, and
  Marques]{schmidt2021crystal}
Jonathan Schmidt, Love Pettersson, Claudio Verdozzi, Silvana Botti, and
  Miguel~AL Marques.
\newblock Crystal graph attention networks for the prediction of stable
  materials.
\newblock \emph{Science Advances}, 7\penalty0 (49):\penalty0 eabi7948, 2021.

\bibitem[Sch{\"u}tt et~al.(2017)Sch{\"u}tt, Kindermans, Sauceda~Felix, Chmiela,
  Tkatchenko, and M{\"u}ller]{schutt2017schnet}
Kristof Sch{\"u}tt, Pieter-Jan Kindermans, Huziel~Enoc Sauceda~Felix, Stefan
  Chmiela, Alexandre Tkatchenko, and Klaus-Robert M{\"u}ller.
\newblock Schnet: A continuous-filter convolutional neural network for modeling
  quantum interactions.
\newblock \emph{Advances in neural information processing systems}, 30, 2017.

\bibitem[Seko et~al.(2015)Seko, Togo, Hayashi, Tsuda, Chaput, and
  Tanaka]{seko2015prediction}
Atsuto Seko, Atsushi Togo, Hiroyuki Hayashi, Koji Tsuda, Laurent Chaput, and
  Isao Tanaka.
\newblock Prediction of low-thermal-conductivity compounds with
  first-principles anharmonic lattice-dynamics calculations and bayesian
  optimization.
\newblock \emph{Physical review letters}, 115\penalty0 (20):\penalty0 205901,
  2015.

\bibitem[Seko et~al.(2017)Seko, Hayashi, Nakayama, Takahashi, and
  Tanaka]{seko2017representation}
Atsuto Seko, Hiroyuki Hayashi, Keita Nakayama, Akira Takahashi, and Isao
  Tanaka.
\newblock Representation of compounds for machine-learning prediction of
  physical properties.
\newblock \emph{Physical Review B}, 95\penalty0 (14):\penalty0 144110, 2017.

\bibitem[Smith(2018)]{smith2018disciplined}
Leslie~N Smith.
\newblock A disciplined approach to neural network hyper-parameters: Part
  1--learning rate, batch size, momentum, and weight decay.
\newblock \emph{arXiv preprint arXiv:1803.09820}, 2018.

\bibitem[Vig(2019)]{vig-2019-multiscale}
Jesse Vig.
\newblock A multiscale visualization of attention in the transformer model.
\newblock In \emph{Proceedings of the 57th Annual Meeting of the Association
  for Computational Linguistics: System Demonstrations}, pages 37--42,
  Florence, Italy, July 2019. Association for Computational Linguistics.
\newblock \doi{10.18653/v1/P19-3007}.
\newblock URL \url{https://www.aclweb.org/anthology/P19-3007}.

\bibitem[Ward et~al.(2017)Ward, Liu, Krishna, Hegde, Agrawal, Choudhary, and
  Wolverton]{ward2017including}
Logan Ward, Ruoqian Liu, Amar Krishna, Vinay~I Hegde, Ankit Agrawal, Alok
  Choudhary, and Chris Wolverton.
\newblock Including crystal structure attributes in machine learning models of
  formation energies via voronoi tessellations.
\newblock \emph{Physical Review B}, 96\penalty0 (2):\penalty0 024104, 2017.

\bibitem[Xie and Grossman(2018)]{xie2018crystal}
Tian Xie and Jeffrey~C Grossman.
\newblock Crystal graph convolutional neural networks for an accurate and
  interpretable prediction of material properties.
\newblock \emph{Phys. Rev. Lett.}, 120\penalty0 (14):\penalty0 145301, 2018.

\bibitem[Xie et~al.(2021)Xie, Fu, Ganea, Barzilay, and
  Jaakkola]{xie2021crystal}
Tian Xie, Xiang Fu, Octavian-Eugen Ganea, Regina Barzilay, and Tommi Jaakkola.
\newblock Crystal diffusion variational autoencoder for periodic material
  generation.
\newblock \emph{arXiv preprint arXiv:2110.06197}, 2021.

\bibitem[Yadati et~al.(2019)Yadati, Nimishakavi, Yadav, Nitin, Louis, and
  Talukdar]{yadati2019hypergcn}
Naganand Yadati, Madhav Nimishakavi, Prateek Yadav, Vikram Nitin, Anand Louis,
  and Partha Talukdar.
\newblock Hypergcn: A new method for training graph convolutional networks on
  hypergraphs.
\newblock \emph{Advances in neural information processing systems}, 32, 2019.

\bibitem[Yan et~al.(2022)Yan, Liu, Lin, and Ji]{yan2022periodic}
Keqiang Yan, Yi~Liu, Yuchao Lin, and Shuiwang Ji.
\newblock Periodic graph transformers for crystal material property prediction.
\newblock In \emph{The 36th Annual Conference on Neural Information Processing
  Systems}, 2022.

\bibitem[Ying et~al.(2018)Ying, He, Chen, Eksombatchai, Hamilton, and
  Leskovec]{ying2018graph}
Rex Ying, Ruining He, Kaifeng Chen, Pong Eksombatchai, William~L Hamilton, and
  Jure Leskovec.
\newblock Graph convolutional neural networks for web-scale recommender
  systems.
\newblock In \emph{Proceedings of the 24th ACM SIGKDD international conference
  on knowledge discovery \& data mining}, pages 974--983, 2018.

\end{thebibliography}
\title{CrysMMNet -  Supplementary Materials}
\onecolumn 
\maketitle
\appendix
\section{Qualitative analysis of the attention layers}
\label{supp}
\begin{figure}
	\centering
        \subfloat{\includegraphics[scale=0.8]{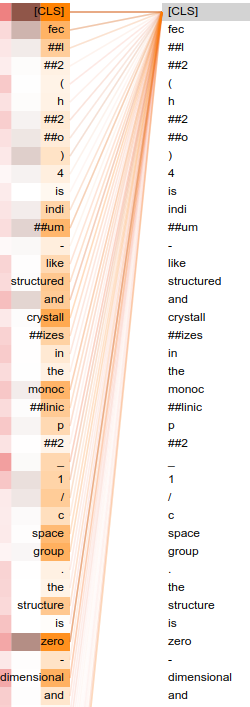}}
	\subfloat{\includegraphics[scale=0.8]{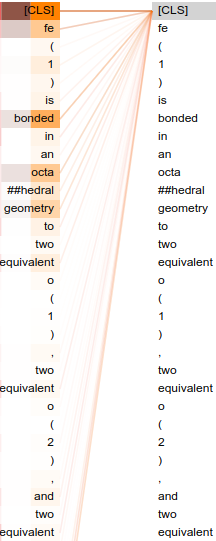}}
	\caption{Attention weights of 5th Layer at MatSciBert between [CLS] token and other tokens in material's description for $\mathbf{FeH_8(ClO_2)_2}$.}
	\label{fig:attn_analysis_1}
\end{figure}

\begin{figure}
	\centering
         \subfloat{\includegraphics[scale=0.8]{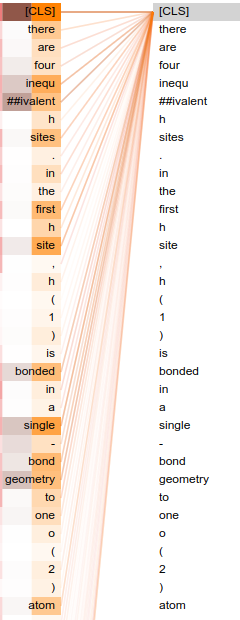}}
         \subfloat{\includegraphics[scale=0.8]{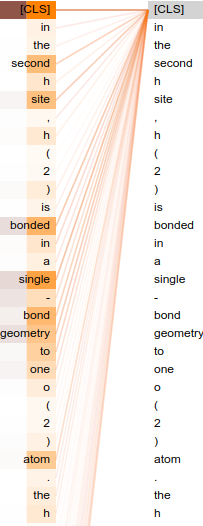}}
	\caption{Attention weights of 5th Layer at MatSciBert between [CLS] token and other tokens in material's description for $\mathbf{FeH_8(ClO_2)_2}$.}
	\label{fig:attn_analysis_2}
\end{figure}

uai2023-supplimentary
In this section, we aim to visualize and understand attention in different tokens in the material description and we perform a qualitative analysis of the attention layer in MatSciBert.  We utilized the standard BertViz tool \footnote{https://github.com/jessevig/bertviz} to analyze and visualize the attention scores in the MatSciBert Model. We present a case study of  the textual data of $FeH_8(ClO_2)_2$ in figure \ref{fig:attn_analysis_1} and \ref{fig:attn_analysis_2}, where we have examined the attention score of the [CLS] token at the 5th layer of MatSciBert. From figure \ref{fig:attn_analysis_1} and \ref{fig:attn_analysis_2}, it is  clearly observed,  MatSciBert allocates higher attention scores to tokens that define global features of the crystal, such as   `Formula', `Mineral', `Crystal System', `Space Group Number', and `Dimensionality'. Further, we observe for local information corresponding to Fe, H, and O atoms. MatScibert provides more attention score to tokens related to bond types (octahedral geometry, equivalent bond, distorted water-like geometry etc) and bond lengths (2.08 Å, 2.10 Å and 2.53 Å bond length). It is evident from these observations that MatSciBert is attending the important tokens related to global and local material information, to generate more expressive multimodal representation.

\end{document}